%% file: source.tex
\documentclass{aa}

\usepackage{graphicx,natbib}
\usepackage{amsmath} 
\usepackage{deluxetable}
\bibpunct{(}{)}{;}{a}{}{,}

\def\ergscm{erg~s$^{-1}$~cm$^{-2}$}

\def\arcmin{\hbox{$^\prime$}}

\def\flux{erg s$^{-1}$ cm$^{-2}$}

\def\apj{ApJ}

\def\aap{A\&A}
\def\mnras{MNRAS}
\newcommand{\gtrsim}{\mathrel{\hbox{\rlap{\hbox{\lower4pt\hbox{$\sim$}}}\hbox{$>$}}}}

\def\agntot{214 }

\def\AgnTotExtraNonblazar{153 }% 4 Blazars at |b|>5

\begin{document}

\title{INTEGRAL/IBIS 7-year All-Sky Hard X-Ray Survey\thanks{Based on observations with INTEGRAL, an ESA project with
instruments and science data centre funded by ESA member states
(especially the PI countries: Denmark, France, Germany, Italy,
Switzerland, Spain), Czech Republic and Poland, and with the
participation of Russia and the USA}\\ Part II: Catalog of Sources.}

\author{R.~Krivonos\inst{1,2}, S.~Tsygankov\inst{1,2},
  M.~Revnivtsev\inst{2,3}, S. Grebenev\inst{2}, E.~Churazov\inst{1,2}
  \and R.~Sunyaev\inst{1,2}}

%\author{R. Krivonos\inst{1,2}, E. Churazov\inst{1,2}, R. Sunyaev\inst{1,2}}
%\author{authors\inst{1,2}}

\institute{
              Max-Planck-Institute f\"ur Astrophysik,
              Karl-Schwarzschild-Str. 1, D-85740 Garching bei M\"unchen,
              Germany
\and
              Space Research Institute, Russian Academy of Sciences,
              Profsoyuznaya 84/32, 117997 Moscow, Russia
\and 
              Excellence Universe Cluster, Munich Technical University, Boltzman-Str. 2, D-85748 Garching bei M\"unchen, Germany
            }
\authorrunning{Krivonos et al.}

\abstract{This paper is the second in a series devoted to the hard
  X-ray ($17-60$~keV) whole sky survey performed by the INTEGRAL
  observatory over seven years. Here we present a catalog of detected
  sources which includes 521 objects, 449 of which exceed a $5\sigma$
  detection threshold on the time-averaged map of the sky, and 53 were
  detected in various subsamples of exposures. Among the identified
  sources with known and suspected nature, 262 are Galactic (101
  low-mass X-ray binaries, 95 high-mass X-ray binaries, 36 cataclysmic
  variables, and 30 of other types) and 219 are extragalactic,
  including \agntot active galactic nuclei (AGNs), 4 galaxy clusters,
  and galaxy ESO~389-G~002. The extragalactic ($|b|>5^{\circ}$) and
  Galactic ($|b|<5^{\circ}$) persistently detected source samples are
  of high identification completeness (respectively $\sim96\%$ and
  $\sim94\%$) and valuable for population studies.   } \maketitle

\section{Introduction}

The INTEGRAL observatory \citep{integral} has been successfully
operating in orbit since its launch in 2002. Due to the high
sensitivity and relatively good angular resolution of its instruments,
in particular the coded-mask telescope IBIS \citep{ibis}, surveying
the sky in hard X-rays is one of the primary goals of INTEGRAL.  The
main scientific results and source catalogues have been reported in
many relevant
papers \citep{revetal03a,molkov2004,krietal05,revetal06,birdI,birdII,birdIII,birdIV,bassani06,bazzano06,krietal07b,sazonov07,beckmann09}.

Recently, great progress in surveying the hard X-ray sky was achieved
with the Burst Alert Telescope \citep[BAT;][]{bat} at the \textit{Swift}
observatory \citep{swift}. As seen from the large sample of detected
Active Galactic Nuclei \citep{bat22,palermo36}, the results of the
\textit{Swift}/BAT all-sky survey are very valuable for extragalactic
studies. 

Contrary to the \textit{Swift}, with a nearly uniform survey, the
INTEGRAL observatory provides the sky survey with exposure more
concentrated in the Galactic Plane (GP). This fact makes the
\textit{Swift}/BAT and INTEGRAL/IBIS surveys complementary to each
other.

In our first paper in a series (Krivonos et al., 2010a, in prep.), we presented
the hard X-ray survey based on the improved sky reconstruction method
for the IBIS telescope. The sensitivity of the survey was
significantly improved due to suppression of the systematic noise. 

Here we present the catalog of sources detected in the survey.

\section{Survey}
\label{section:survey}

With the 7-year mission data (December 2002 -- July 2009) we conducted
the all-sky survey in the working energy band $17-60$~keV. The full
analyzed data set comprises $\sim83$~Ms of effective (dead time-corrected)
exposure. The minimum sensitivity of the survey was
$3.7\times10^{-12}$~\ergscm ~($\sim0.26$~mCrab\footnote{A flux of
  $1$~mCrab in the $17-60$~keV energy band corresponds to
  $1.43\times10^{-11}$ \flux ~for a source with a Crab-like spectrum.}
in 17-60~keV) at a $5\sigma$ detection level. The survey covered
$90\%$ of the sky down to the flux limit of
$6.2\times10^{-11}$~\ergscm ~($\sim4.32$~mCrab) and $10\%$ of the sky
area down to the flux limit of $8.6\times10^{-12}$~\ergscm
~($\sim0.60$~mCrab).

In the current survey we perform a census of hard X-ray sources
detected on the all-time averaged sky frame. However, a number of
sources was detected in a various subsamples of exposures during
periods of outburst activity. Apart from the catalogue, we provided
also the light curves of detected sources averaged over each
spacecraft orbit (3 days). However, we did not attempt to look for
sources on time scales intermediate between one orbit and seven
years. This issue was addressed in the recent catalogue survey by
\cite{birdIV}.

We divided all sources detected in the current survey into the two
classes according to their detection condition. The \textit{Long-term
  Detected (LtD) sources} were found on the 7-year time-averaged map
above $5\sigma$ detection threshold. We checked, that the measured
flux was not dominated by a single event of strong outburst activity,
however the time-averaged flux may contain intrinsic source
variability (Fig.~\ref{fig:example}). The list of \textit{Short-term
  Detected (StD) sources} contains objects significantly detected on
the time scales of spacecraft orbit ($\sim3$~days), or set of orbits
($\sim$~weeks). During 7 years of the INTEGRAL survey, some sources
demonstrated period of strong outburst activity, while over the remaining
time span of observations they were not detected (e.g. 4U~1901+03,
Fig.~\ref{fig:example}). The source in outburst can be so bright, that
it may be detected on the all-time averaged sky map. Nevertheless we
consider these sources as short-term detected. 

The above classification did not strictly follow the physical
understanding of persistent and transient sources. Some objects
(except one-time events) may move from \textit{LtD} to \textit{StD}
and vice versa with the new observational data and other selection
criteria. The exact classification of sources we leave for
the interested reader. To do this, we provide light curves of detected
sources and histograms of its flux distribution
(Sect.\ref{section:conclusion}). As a demonstration, we show two
examples listed in the catalogue, the \textit{LtD} source LMXB
GX~349+2, and HMXB transient 4U~1901+03 as \textit{StD} source in
outburst \citep{galloway05}.

With a new data sets obtained by the INTEGRAL since 2006, a
number of faint sources with a known nature detected in our
previous survey \citep[][referred to as K07]{krietal07b} fell below a
$5\sigma$ detection threshold, probably due to intrinsic
variability. We included $19$ known catalogued sources in the current
survey with detection significance in the range $4.7-5.0\sigma$. However we
emphasize that for statistical studies only those \textit{LtD}
(persistent) sources should be used from the catalog that have
statistical significance higher than $5\sigma$.

\section{Detection of sources}
\label{section:survey:srcdetect}

We performed a search for sources on $25^\circ\times25^\circ$ sky
mosaics covering the whole sky. By analogy with K07 the sources were
searched as excesses on ISGRI sky maps, convolved with a Gaussian
representing the effective instrumental PSF.

The search was made on minimum time scale of each spacecraft orbit (3
days) and the whole time span of 7 years.  Following to K07 we adopted
the corresponding detection thresholds of $(S/N)_{\rm lim}>5.5\sigma$
and $(S/N)_{\rm lim}>5\sigma$ to ensure that the final catalog
contains less than 1--2 spurious sources.

By searching the final average map for the local maxima, we found 449
excesses above $5\sigma$. The list of transiently detected sources
contains $53$ objects. The positions of newly detected sources were
cross-correlated with SIMBAD and NED catalogues using a $4.2$~arcmin
search radius ($90\%$ confidence level for a source detected at 5-6
standard deviations, K07), and the recent \textit{Swift} survey source
catalogues reported in papers by \cite{bat22} and
\cite{palermo36}. Utilizing the whole available information for the
sources with firm identification and sources with tentative but
unconfirmed classification of a given type (later refered as having
``a suspected origin''), we have identified 219 extragalactic objects
and 262 galactic sources.  The total number of unidentified sources on
the time averaged map above $5\sigma$ detection threshold is $43$.
Most of them ($31$) are located in the Galactic Plane at latitudes
$|b|<5^{\circ}$ (see Table~\ref{table:stat} for source statistics).

\section{Catalog of sources}
\label{sec:catalog}

The full list of sources is presented in Table \ref{tab:catalog}, and its content is
described below.

{\it Column (1) ``Id''} -- source sequence number in the catalog.

{\it Column (2) ``Name''} -- source name. For sources whose nature was known 
before their detection by INTEGRAL, their common names are given. Sources
discovered by INTEGRAL or those whose nature was established thanks to
INTEGRAL are named ``IGR''

{\it Columns (3,4) ``RA, Dec''} -- source Equatorial (J2000) coordinates.

{\it Column (5) ``Flux, $17-60$~keV''} -- time-averaged source flux in mCrab units.

{\it Column (6) ``Type''} -- general astrophysical type of the object: LMXB
(HMXB) -- low- (high-) mass X-ray binary, AGN -- active galactic
nucleus, SNR/PWN -- supernova remnant, CV -- cataclysmic variable, PSR
-- isolated pulsar or pulsar wind nebula, SGR -- soft gamma repeater,
RS CVn -- coronally active binary star, SymbStar -- symbiotic star,
Cluster -- cluster of galaxies. The question mark indicates
  that the specified type is not firmly determined, and should be
  confirmed. The census of these sources is marked in
  Table~\ref{table:stat} with $S$ index.

{\it Column (7) ``Ref.''} -- references. These are mainly provided for new
sources and are related to their discovery and/or nature.

{\it Column (8) ``Notes''} -- additional notes like type subclass, redshift
information, alternative source names, etc. Redshift of the
extragalactic sources was obtained from the SIMBAD and NED database.

In Table~\ref{table:stat} we presented source statistics for
types, detections in Galactic Plane ($|b|<5^{\circ}$), high galactic latitude sky ($|b|>5^{\circ}$),
and comparison with our previous 4-year survey K07.

\begin{table*}
\begin{center}
\caption{ Catalog source statistics and comparison with the previous
  survey K07. The star symbol denotes number of sources with detection
  threshold above $5\sigma$. The number of sources with tentative
  classification of a known type is denoted with $S$ index
  (``suspected''). Suspected identifications are distributed over the
  categories behind the + sign in addition to the secure ones, but counted among
  NotID. All sources with suspected nature are above $5\sigma$
  detection threshold.}
\label{table:stat}
%\begin{flushleft}
{\normalsize
\begin{tabular}{lccccccc}
\hline 
\hline 
\noalign{\smallskip}   & AGN & LMXB & HMXB & CV & Other &  NotID & Total \\ 
\noalign{\smallskip}\hline\noalign{\smallskip} \hline
\multicolumn{7}{c}{Current work -- over 7 years}  \\
\textit{StD} & 4$^{*}$+1$^{s}$        & 16$^{*}$+2$^{s}$ & 15$^{*}$+1$^{s}$   & 2$^{*}$     & 1$^{*}$+1$^s$  & $15(15^{*})-5^{s}$ &  53$^{*}$ \\
\textit{LtD} & 202(190$^{*}$)+7$^{s}$ & 82$^{*}$+1$^{s}$ & 70$^{*}$+9$^{s}$   & 32(31$^*$)+2$^s$  & 33$^{*}$ & $49(43^{*})-19^{s}$ &  468(449$^{*}$) \\
         All & 206(194$^{*}$)+8$^{s}$ & 98$^*$+3$^{s}$   & 85$^*$+10$^{s}$    & 34(33$^*$)+2$^s$  & 34$^{*}$+1$^s$ & $64(58^*)-24^s$ &  521(502$^*$) \\

\hline
\multicolumn{7}{c}{Galactic latitude selection $|b|<5^{\circ}$}  \\
\textit{StD} & --                 & 14$^{*}$+1$^{s}$ & 15$^{*}$+2$^{s}$  & --     &  1$^{s}$    & $11(11^{*})-4^{s}$       &   40$^{*}$ \\
\textit{LtD} & 32$^{*}$+4$^{s}$   & 59$^{*}$+1$^{s}$ & 64$^{*}$+9$^{s}$  & 12$^*$+2$^{s}$ & 25$^*$ & 34(31$^{*})-16^{s}$ &  244(241$^{*}$) \\
         All & 32$^{*}$+4$^{s}$   & 73$^*$+3$^{s}$   & 79$^*$+10$^{s}$   & 12$^*$+2$^{s}$ & 25$^*$+1$^{s}$ & $45(42^*)-20^s$     & 266(263$^*$)  \\

\hline
\multicolumn{7}{c}{Galactic latitude selection $|b|>5^{\circ}$}  \\
\textit{StD} & 4$^{*}$+1$^{s}$        & 2$^{*}$  & --      & 2$^*$      & 1$^*$ & $4(4^{*})-1^{s}$ &  13$^{*}$ \\
\textit{LtD} & 170(158$^{*}$)+3$^{s}$ & 23$^{*}$ & 6$^*$ & 20(19$^*$) & 8$^*$ & $15(12^{*})-3^{s}$ &  242(226$^{*}$) \\
         All & 174(162$^{*}$)+4$^{s}$ & 25$^*$   & 6$^*$   & 22(21$^*$) & 9$^*$ & $19(16^*)-4^s$ & 255(239$^*$)  \\

\hline
\hline
\multicolumn{7}{c}{K07 -- over 4 years}  \\
\textit{StD} & 1+1$^s$               & 7$^*$+1$^s$          & 3$^*$            & 2(1$^*$) & 1$^*$+1$^s$       & $14(14^*)-3^s$ &  28(26$^*$) \\
\textit{LtD} & 129(92$^{*}$)+2$^s$   & 77(76$^*$)+5$^s$     & 69(66$^*$)+4$^s$  & 19(13$^*$) & 29(26$^*)+4^s$ & $52(43^*)-15^s$ &  375(316$^*$) \\
All        & 130(92$^*$)+3$^s$     & 84(83$^*$)+6$^s$     & 72(69$^*$)+4$^s$  & 21(14$^*$) & $30(27^*)+5^s$ & $66(57^*)-18^s$ &  403(342$^*$) \\
\hline
\noalign{\smallskip}\hline\noalign{\smallskip}
\end{tabular}
}
%\end{flushleft}
\end{center}
\end{table*}

{\it Active Galactic Nuclei} -- the AGN sample was substantially
increased by a factor of 2 with respect to the K07 due to increased
extragalactic exposure. Most of the objects were detected on the
7-year time-averaged sky. About thirty AGNs were found in the Galactic
Plane. The statistically clear sample of 158 AGNs, confidently
detected ($>5\sigma$) and selected in the extragalactic sky
($|b|>5^{\circ}$), is very valuable for the AGN population studies
because of high identification completeness of the survey, which is
$(N_{Tot}-N_{NotID})/N_{Tot}=1-12/226=0.95$. Moreover, taking into
account three tentative AGN classifications, the survey's
identification completness at $|b|>5^{\circ}$ becomes slightly higher
($0.96$).

{\it LMXB and HMXB} -- the low- and high- mass X-ray binaries, as
before, dominate the Galactic sample of the survey. As seen in
Table~\ref{table:stat}, the number of LMXBs and HMXBs, was increased
mainly by short-term detected sources. We should note here,
that with a new observational data, 13 HMXBs and 6 LMXBs persistently
detected in K07 were moved now to the \textit{StD}s\footnote{{\scriptsize
    \textit{HMXB:} V~0332+53, A~0535+262, IGR~J21343+4738, 4U~0115+63,
    IGR~J16358-4726, GRO~J1008-57, IGR~J11215-5952, XTE~J1543-568,
    IGR~J16465-4507, KS~1716-389, A~1845-024, XTE~J1858+034,
    4U~1901+03; \textit{LMXB:} IGR~J00291+5934, XTE~J1550-564,
    XTE~J1720-318, SLX~1746-331, XTE~J1807-294, XTE~J1817-330.}}
according to detection conditions described in
Sect.~\ref{section:survey}.

{\it Cataclysmic Variables} -- similar to the AGNs sample, the number
of CVs was increased by a factor of 2 thanks to the additional high
galactic latitude observations. Most of the CVs were recorded as
\textit{LtD}, except FO~Aqr and V1062~Tau. The position of FO~Aqr has
very poor coverage by INTEGRAL observations and the source was
significantly detected during only one spacecraft orbit. V1062~Tau is
located in the region with a high systematic noise from the bright
source Crab Nebula which prevented its persistent detection. However,
during the 215~ks observations of Crab in August 2003, the source
V1062~Tau was detected with significance $\sim7\sigma$.

{\it Other types} -- the other populations of sources (Clusters, SNR, PSR,
Symbiotic stars, etc.) were persistently detected on the 7-year maps,
and mainly in the Galactic Plane. The total number was not
substantially changed since K07. The number of Clusters of Galaxies
was increased by detection of Triangulum A Cluster, in addition to
Coma, Perseus, and Oph Cluster.

{\it Unidentified sources} -- dominantly in the Galactic Plane and
mainly \textit{LtD}s. 31 unidentified objects detected above $5\sigma$
threshold at $|b|<5^{\circ}$, made the survey in the GP
  identified at level of $\sim87\%$. If we take into account a
suspected nature of 16 \textit{LtD} sources, the identification
  completness of the survey at $|b|<5^{\circ}$ becomes
$\sim94\%$. Most of the unidentified \textit{transiently} detected
sources (\textit{StD}s) were found in the GP, which implicitly points
to their Galactic and probably X-ray binary origin.

\begin{figure*}[t]
\includegraphics[width=0.5\textwidth,bb=18 375 592 718]{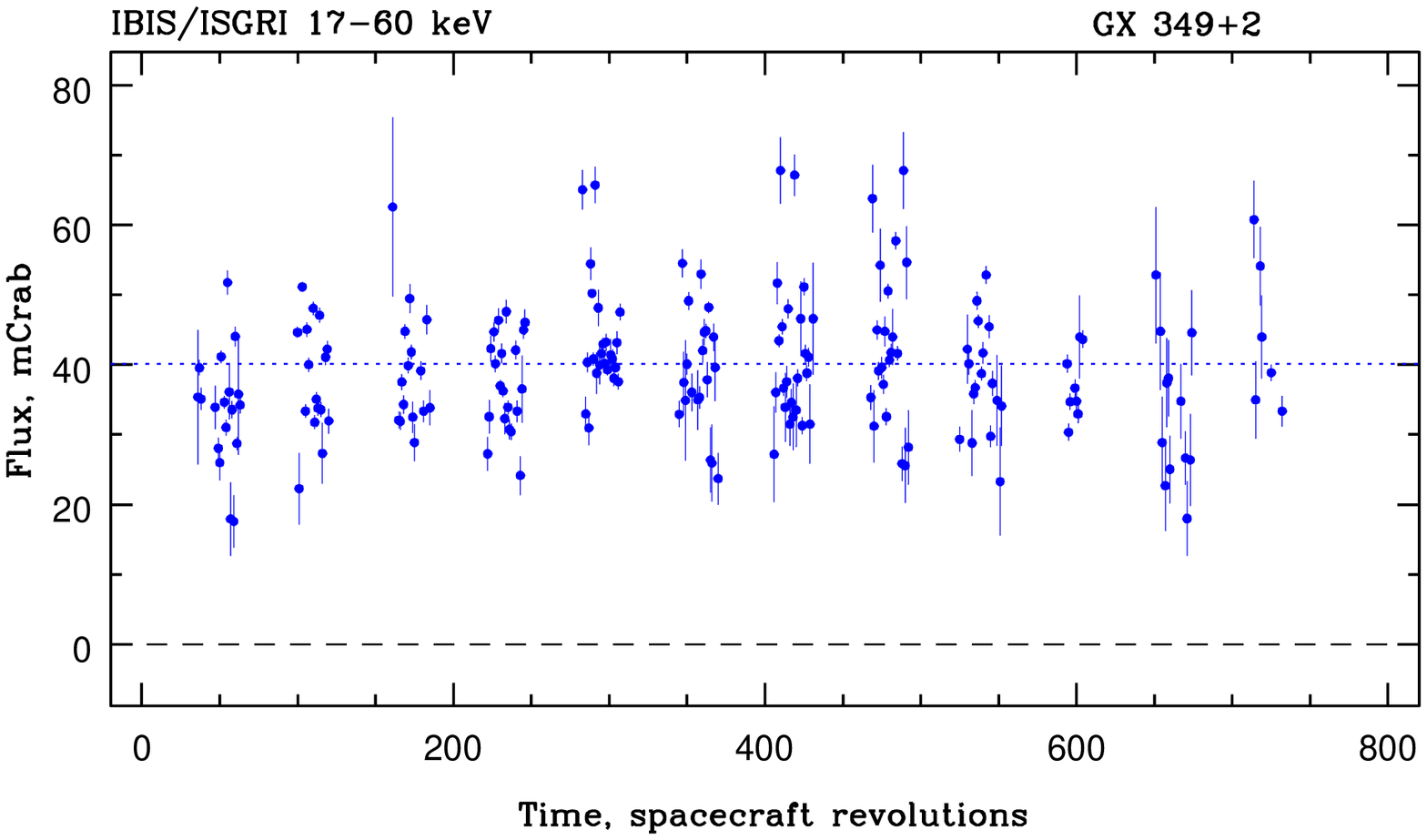}
\includegraphics[width=0.5\textwidth,bb=18 375 592 718]{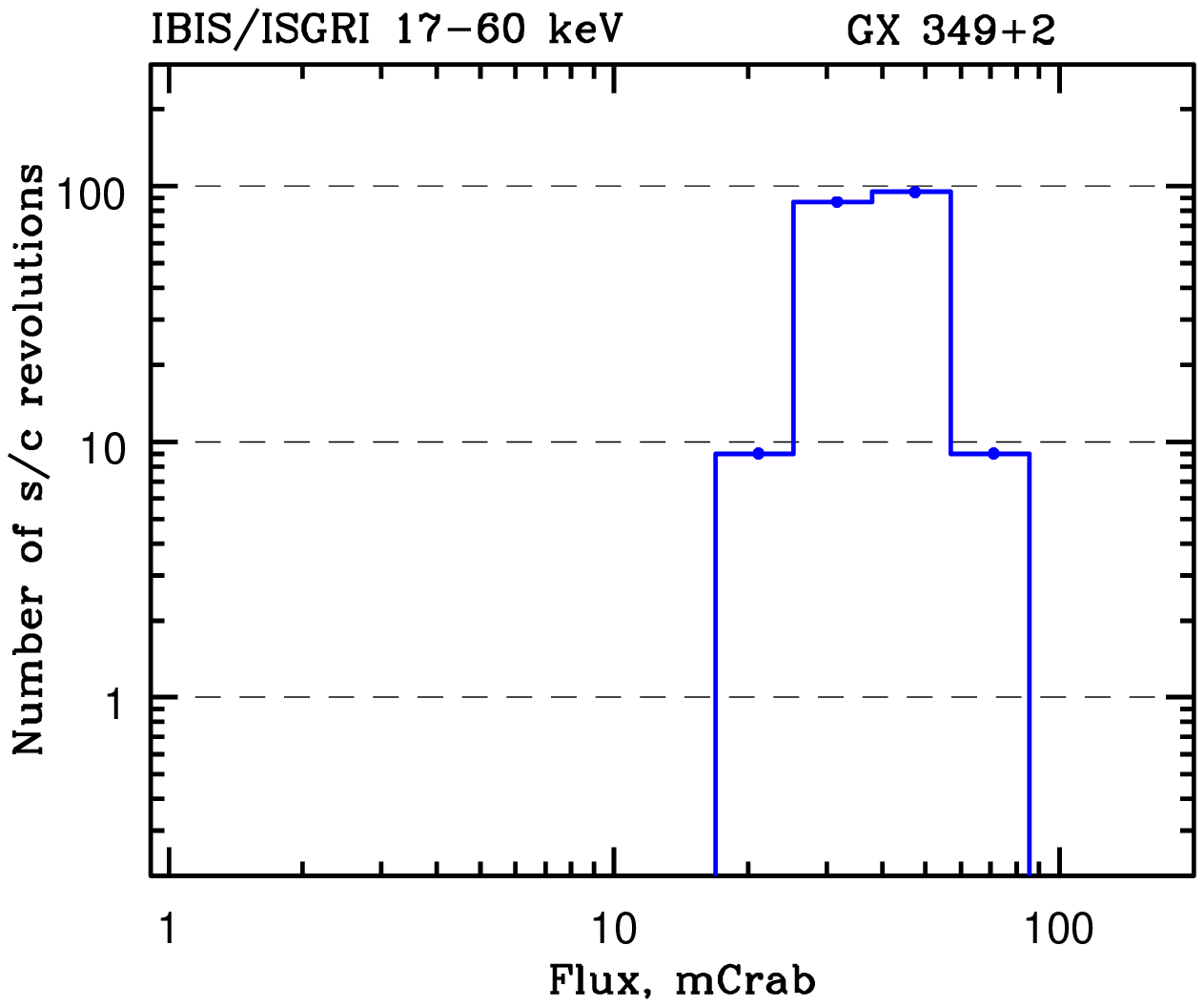}
\includegraphics[width=0.5\textwidth,bb=18 375 592 718]{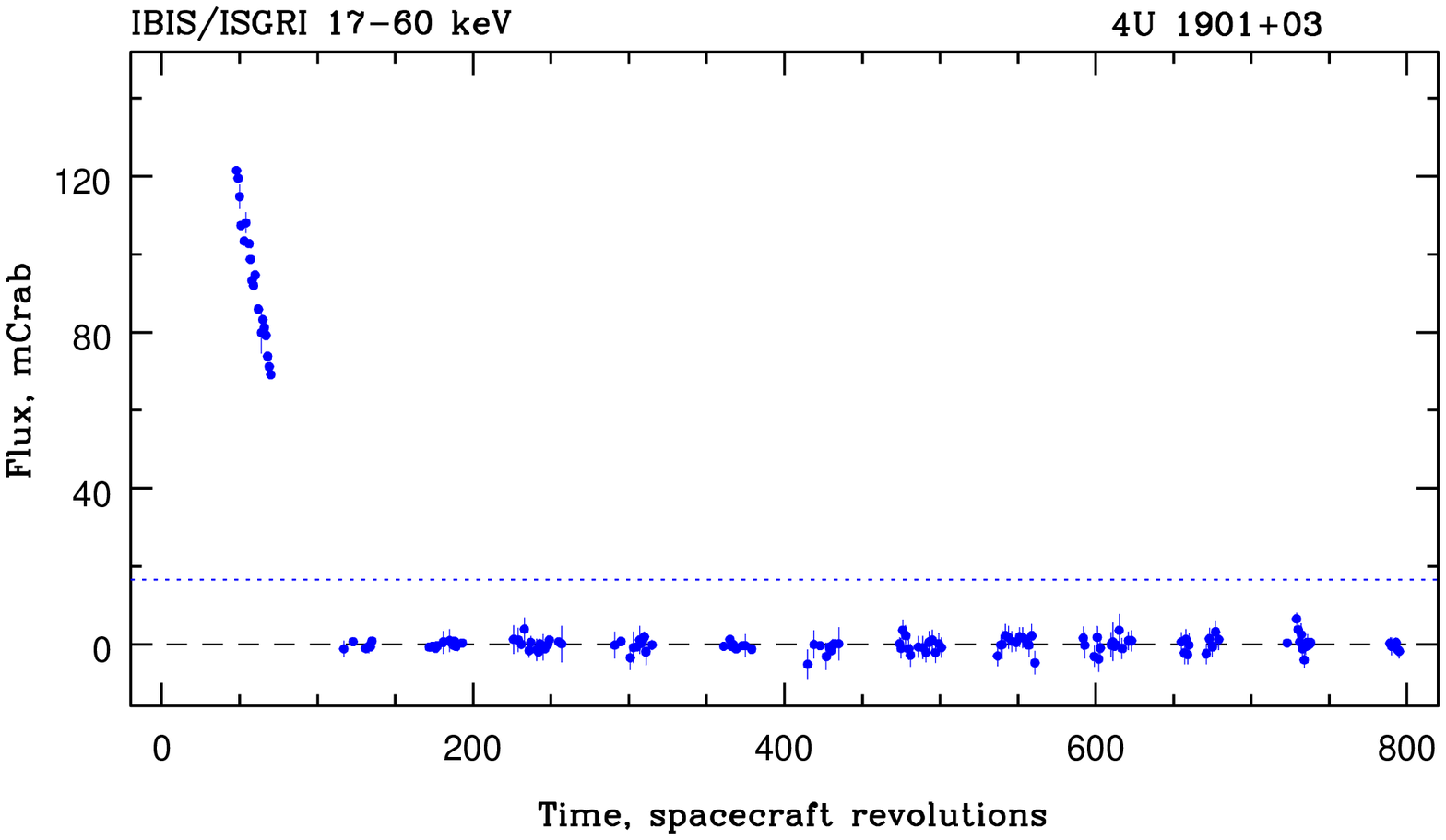}
\includegraphics[width=0.5\textwidth,bb=18 375 592 718]{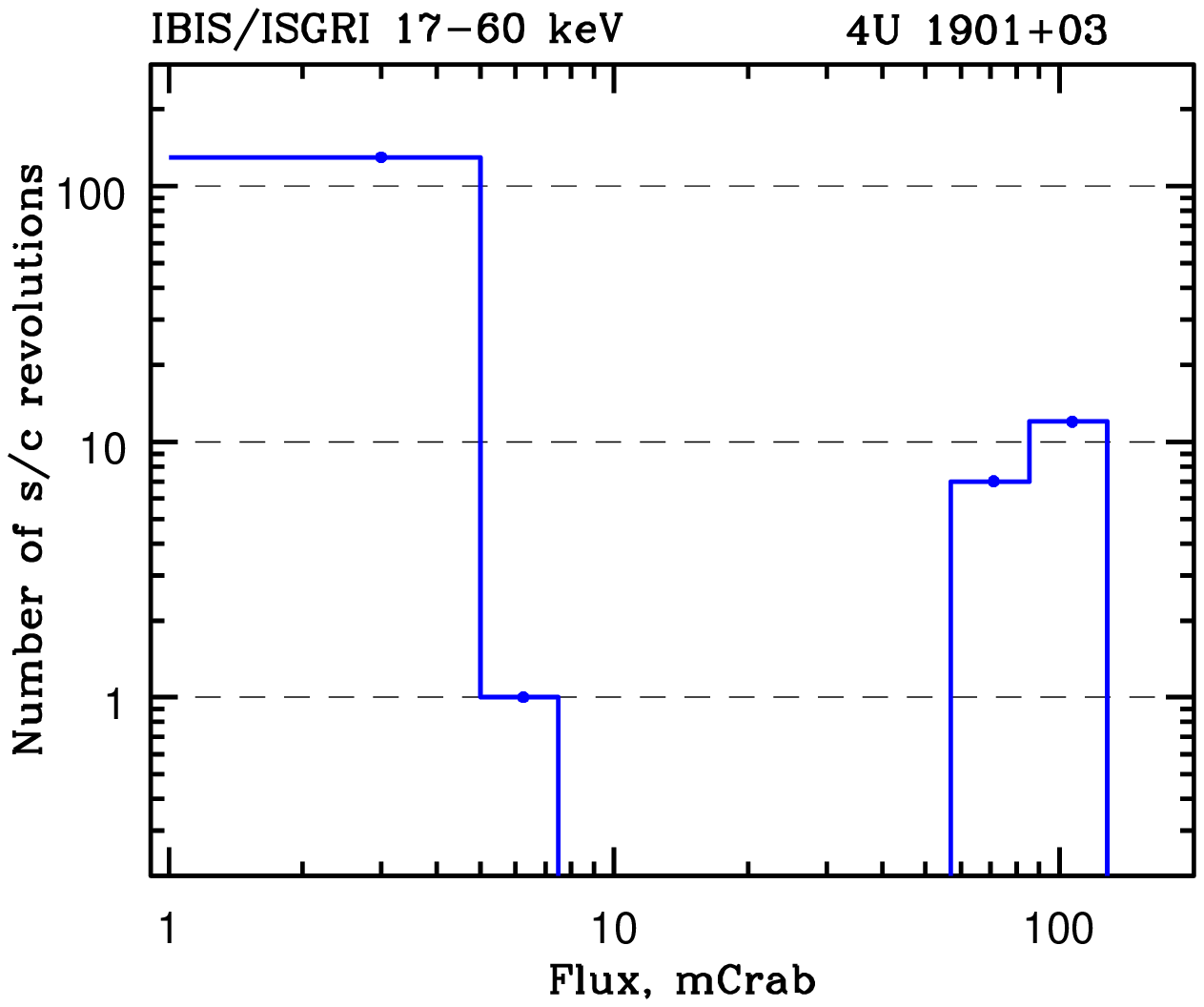}
\caption{The $17-60$~keV light curves (left) and histograms of the
  corresponding flux distribution (right) of two sources in the
  catalog: persistently detected and highly variable LMXB GX~349+2, and
  HMXB transient 4U~1901+03. The blue dotted line in the left figures
  represents flux of the sources measured on 7-year time-averaged
  map. The first flux bin in the right histograms contains
  counts from the range $[-5,5]$~mCrab, and the flux measurements with
  error $>5$~mCrab were dropped.}\label{fig:example}
\end{figure*}

\section{Extragalactic LogN-LogS}
\label{section:survey:lognlogs}

\begin{figure}
 \includegraphics[width=0.5\textwidth]{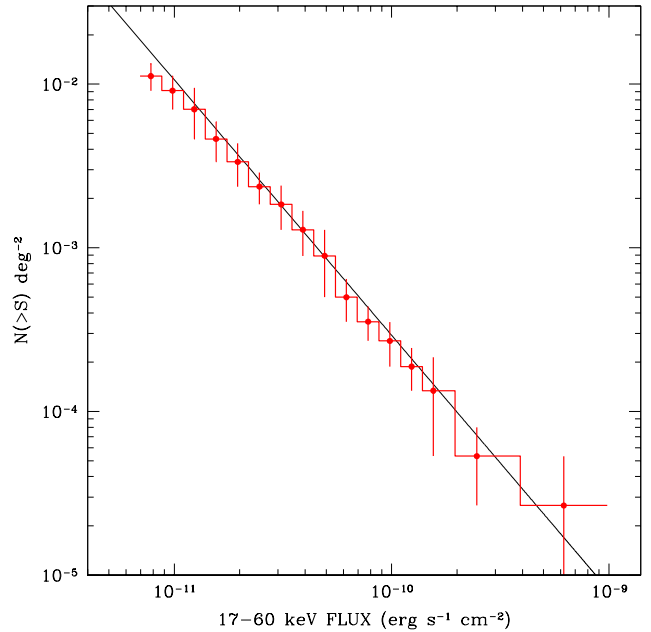}
\caption{Number flux relation of extragalactic objects at
  $|b|>5^{\circ}$ (red points) build from non-blazar AGN sample
  containing \AgnTotExtraNonblazar objects detected above
  $5\sigma$. The best-fitting power law with a slope of $1.56\pm0.10$
  and normalization of $(3.59\pm0.35)\times10^{-3}$ deg$^{-2}$ at
  $S=2\times10^{-11}$ \flux shown by the solid
  line.}\label{fig:lognlogs}
\end{figure}

Under the assumption that AGNs are uniformly distributed over the sky,
we can construct the number-flux function of hard X-ray emitting AGNs.
Since INTEGRAL observations cover the sky inhomogeneously, we should
take the sensitivity map into account in constructing number-flux
functions. {\bf This was done by dividing the source counts by the sky
coverage at the $5\sigma$ level as a function of flux (see Fig.~12 in
Krivonos et al., 2010a, in prep.).} In Fig.~\ref{fig:lognlogs} we show
the cumulative $\log N$--$\log S$ distribution of
\AgnTotExtraNonblazar non-blazar AGNs derived over the whole sky
excluding the Galactic Plane ($|b|<5^{\circ}$). The $\log N$--$\log S$
distribution can be fitted well by a power law: $N(>S)=A
S^{-\alpha}$. Using a maximum-likelihood estimator \citep[see
  e.g.][]{jauncey67,crawford1970}, we determined the best-fit values
of the slope and normalization: $\alpha=1.56\pm0.10$ and
$A=(3.59\pm0.35)\times10^{-3}$ deg$^{-2}$ at $S=2\times10^{-11}$
\flux. The observed $\log N$--$\log S$ slope is consistent with a
homogeneous distribution of sources in space ($\alpha= 3/2$), and
implies that AGNs with fluxes exceeding the survey detection threshold
at the extragalactic coverage ($|b|>5^{\circ}$) account for $\sim1\%$
of the intensity of the cosmic X-ray background in the 17--60~keV
band.

%%%%%%%%%%%%%%%%%%%%%%%%%%%%%%%%%%%%%%%%%%%%%%%%%%%%%%%%%%%%%%%%
%%% $A=(6.06\pm0.63)\times10^{-3}$ deg$^{-2}$ at $S=1$~mCrab %%%
%%%%%%%%%%%%%%%%%%%%%%%%%%%%%%%%%%%%%%%%%%%%%%%%%%%%%%%%%%%%%%%%

\section{Concluding remarks}
\label{section:conclusion}

We presented the catalogue of sources detected in the hard X-ray
(17--60~keV) whole sky survey performed at the INTEGRAL observatory
over seven years (Krivonos et al., 2010a, in press).

Our catalog contains 521 sources of different types. According to
detection conditions, we divided all sources on long-term and
short-term detected (\textit{LtD} and \textit{StD}). The statistically
clear sample of 449 \textit{LtD} sources was found on the averaged sky
map above $5\sigma$ detection level. 53 \textit{StD} sources were
detected in the different subsamples of exposures.

Among Galactic sources with firmly known and suspected nature, we
found 101 LMXBs, 95 HMXBs, 36 CVs, and 30 of other types. Among known
and suspected extragalactic identifications, we found 213 AGNs, and
the rest is Galaxy Clusters (4) and galaxy ESO~389-G~002. We presented
the detailed catalog source statistics in the Table~\ref{table:stat}.

We would like to stress that our survey is of high identification
completeness with respect to the confidently detected ($>5\sigma$) and
persistent (\emph{LtD}) sources. Taking into account detected objects
with firm and tentative classification, the survey's completeness in
Galactic Plane ($|b|<5^{\circ}$) and extragalactic selection
($|b|>5^{\circ}$) is $\sim94\%$ and $\sim96\%$, respectively.

Our survey provides the highest sensitivity in the Galactic Plane,
reaching the limiting flux of $\sim0.26$~mCrab or
$3.7\times10^{-12}$~\ergscm ~in the working energy band $17-60$~keV.
The faintest Galactic source is a type-I X-ray burster AX~J1754.2-2754
\citep{2007ATel.1094....1C,2007AstL...33..807C} detected on the
time-averaged map at $6.4\sigma$ with a flux of $0.32$~mCrab
($4.6\times10^{-12}$~\ergscm).

The Galactic sample of the new survey allows us to significantly
extend the study of the faint end of galactic X-ray binaries
population \citep{revetal08} with luminosities $\sim
4\times10^{34}~{erg~s^{-1}}$ (at the distance of the Galactic Center).

Apart from the catalogue of sources available
online\footnote{http://hea.iki.rssi.ru/integral}$^,$\footnote{http://www.mpa-garching.mpg.de/integral},
we provide to the scientific community the light curves of detected
sources averaged over each INTEGRAL orbit (3~days) and histograms of
the corresponding flux distribution (see examples in
Fig.\ref{fig:example}).

\begin{acknowledgements}
The data used were obtained from the European and Russian INTEGRAL
Science Data
Centers\footnote{http://isdc.unige.ch}$^,$\footnote{http://hea.iki.rssi.ru/rsdc}.
The work was supported by the President of the Russian Federation
(through the program supporting leading scientific schools, project
NSH-5069.2010.2), by the Presidium of the Russian Academy of
Sciences/RAS (the program P19 ``Origin, Structure, and Evolution of
Objects of the Universe''), by the Division of Physical Sciences of
the RAS (the program ``Extended objects in the Universe'', OFN-16).
\end{acknowledgements}

\section{References for the catalog}
\label{section:references}
\footnotesize
{\bf [1]} \cite{2005A&A...433.1163D},
{\bf [2]} \cite{2005A&A...444L..37S},
{\bf [3]} \cite{2004A&A...426L..41M},
{\bf [4]} \cite{2006A&A...449.1139M},
{\bf [5]} \cite{2005ATel..622....1B},
{\bf [6]} \cite{2003ATel..181....1S},
{\bf [7]} \cite{2006A&A...455...11M},
{\bf [8]} \cite{2006A&A...459...21M},
{\bf [9]} \cite{bassani06},
{\bf [10]} \cite{2006ApJ...646..452S},
{\bf [11]} \cite{2004ATel..329....1L},
{\bf [12]} \cite{2005ATel..469....1L},
{\bf [13]} \cite{2004ATel..261....1D},
{\bf [14]} \cite{2004ATel..275....1G},
{\bf [15]} \cite{2006ApJ...638.1045S},
{\bf [16]} \cite{2004ATel..345....1K},
{\bf [17]} \cite{2005ATel..519....1C},
{\bf [18]} \cite{2004ATel..362....1I},
{\bf [19]} \cite{2004ATel..340....1R},
{\bf [20]} \cite{2006ATel..785....1M},
{\bf [21]} \cite{2004A&A...425L..49R},
{\bf [22]} \cite{2006ATel..874....1K},
{\bf [23]} \cite{revetal06},
{\bf [24]} \cite{2003AstL...29..719L},
{\bf [25]} \cite{2005A&A...433L..41L},
{\bf [26]} \cite{2004A&A...427L..21B},
{\bf [27]} \cite{2005ApJ...634L..21B},
{\bf [29]} \cite{1992SvAL...18...88P},
{\bf [30]} \cite{2003ATel..151....1L},
{\bf [31]} \cite{2005A&A...441L...1I},
{\bf [32]} \cite{2006AstL...32..221B},
{\bf [33]} \cite{2006ATel..831....1N},
{\bf [34]} \cite{2004ATel..278....1P},
{\bf [35]} \cite{2005A&A...437L..27B},
{\bf [36]} \cite{2006MNRAS.372..224B},
{\bf [37]} \cite{2006A&A...453..133W},
{\bf [38]} \cite{molkov2004},
{\bf [39]} \cite{2006ApJ...649L..21B},
{\bf [40]} \cite{2000A&AS..147...25L},
{\bf [41]} \cite{2001A&A...368.1021L},
{\bf [42]} \cite{2005A&A...432..235R},
{\bf [43]} \cite{2005ATel..394....1D},
{\bf [44]} \cite{birdII},
{\bf [45]} \cite{2004ATel..281....1D},
{\bf [46]} \cite{2005ApJ...629L.109U},
{\bf [47]} \cite{2005ApJ...630L.157M},
{\bf [48]} \cite{1995AstL...21..431A},
{\bf [49]} \cite{2006MNRAS.366..918C},
{\bf [50]} \cite{2006AstL...32..588B},
{\bf [51]} \cite{2005MNRAS.364..455C},
{\bf [52]} \cite{2005A&A...440..637R},
{\bf [53]} \cite{2003IAUC.8088....4H},
{\bf [54]} \cite{2003ATel..190....1S},
{\bf [55]} \cite{2006ATel..883....1B},
{\bf [56]} \cite{2005ATel..470....1N},
{\bf [57]} \cite{2006ATel..810....1K},
{\bf [58]} \cite{2005ATel..545....1K},
{\bf [60]} \cite{2006ATel..880....1B},
{\bf [61]} \cite{2005ATel..616....1G},
{\bf [62]} \cite{2006ATel..684....1K},
{\bf [63]} \cite{2006ATel..783....1M},
{\bf [64]} \cite{2004ATel..328....1L},
{\bf [65]} \cite{2005ATel..669....1T},
{\bf [66]} \cite{2006ATel..847....1H},
{\bf [67]} \cite{2006ATel..715....1M},
{\bf [68]} \cite{2005ATel..528....1M},
{\bf [69]} \cite{2007AstL...33..159K},
{\bf [70]} \cite{2006ATel..778....1B},
{\bf [71]} \cite{2003ATel..157....1C},
{\bf [72]} \cite{2005MNRAS.361..141G},
{\bf [73]} \cite{2005ATel..624....1T},
{\bf [74]} \cite{2005ATel..467....1G},
{\bf [75]} \cite{2006ATel..813....1G},
{\bf [76]} \cite{2007A&A...463..957K},
{\bf [77]} \cite{sazonov07},
{\bf [78]} \cite{2006ATel..713....1R},
{\bf [79]} \cite{2006ApJ...636..275B},
{\bf [80]} \cite{2005ATel..546....1P},
{\bf [81]} \cite{2005ATel..551....1T},
{\bf [82]} \cite{2007A&A...467..249S},
{\bf [83]} \cite{2006ApJ...647.1309T},
{\bf [84]} \cite{2003AstL...29..587R},
{\bf [85]} \cite{2003IAUC.8063....3C},
{\bf [86]} \cite{2004ATel..229....1W},
{\bf [87]} \cite{2003IAUC.8076....1T},
{\bf [88]} \cite{2003IAUC.8097....2R},
{\bf [89]} \cite{2004ApJ...602L..45P},
{\bf [90]} \cite{2006A&A...447.1027B},
{\bf [91]} \cite{2004ATel..224....1T},
{\bf [92]} \cite{2003ATel..176....1M},
{\bf [93]} \cite{2005A&A...444..821L},
{\bf [94]} \cite{2003ATel..149....1K},
{\bf [95]} \cite{2003ATel..132....1R},
{\bf [96]} \cite{2003ATel..155....1L},
{\bf [97]} \cite{2005A&A...430..997L},
{\bf [98]} \cite{2004AstL...30..382R},
{\bf [99]} \cite{2003ApJ...596L..63A},
{\bf [100]} \cite{2003ATel..154....1L},
{\bf [101]} \cite{2008A&A...487..509S},
{\bf [102]} \cite{2007ATel.1034....1M},
{\bf [103]} \cite{2005ATel..444....1G},
{\bf [104]} \cite{2004ATel..342....1G},
{\bf [105]} \cite{2005ApJ...631..506B},
{\bf [106]} \cite{2005ATel..457....1G},
{\bf [107]} \cite{2005ATel..654....1K},
{\bf [108]} \cite{2007A&A...470..331M},
{\bf [109]} \cite{2007ApJ...669L...1B},
{\bf [110]} \cite{2006A&A...446L..17B},
{\bf [111]} \cite{2007ATel.1253....1R},
{\bf [112]} \cite{2007ATel.1270....1B},
{\bf [113]} \cite{biketal08},
{\bf [114]} \cite{2007ATel.1245....1K},
{\bf [116]} \cite{2008A&A...482..113M},
{\bf [117]} \cite{2008ATel.1396....1N},
{\bf [118]} \cite{2008ATel.1488....1K},
{\bf [119]} \cite{2007ATel.1319....1G},
{\bf [120]} \cite{2008ApJ...685.1143T},
{\bf [121]} \cite{2006ATel..885....1S},
{\bf [122]} \cite{2008ATel.1620....1M},
{\bf [123]} \cite{2008ATel.1651....1A},
{\bf [124]} \cite{2009A&A...494..417R},
{\bf [125]} \cite{2005ATel..456....1S},
{\bf [126]} \cite{2009MNRAS.395L...1B},
{\bf [127]} \cite{2008ATel.1540....1P},
{\bf [128]} \cite{2008A&A...477L..29L},
{\bf [130]} \cite{2008ATel.1649....1T},
{\bf [131]} \cite{2008ATel.1539....1L},
{\bf [132]} \cite{bat22},
{\bf [134]} \cite{masetti09},
{\bf [135]} \cite{2009A&A...493..339W},
{\bf [136]} \cite{2009ApJ...701..811T},
{\bf [137]} \cite{2009ATel.2132....1M},
{\bf [138]} \cite{2008ATel.1653....1S},
{\bf [139]} \cite{2008A&A...482..731R},
{\bf [140]} \cite{2009ATel.2170....1K},
{\bf [141]} \cite{1999MNRAS.306..100R},
{\bf [142]} \cite{2007ATel.1286....1T},
{\bf [143]} \cite{2008ApJ...674..686W},
{\bf [144]} \cite{2008AstL...34..367B},
{\bf [145]} \cite{2007ApJ...664...64I},
{\bf [146]} \cite{2008AstL...34..653B},
{\bf [147]} \cite{2006ATel..972....1W},
{\bf [148]} \cite{2006ATel..970....1B},
{\bf [149]} \cite{2006ATel..939....1K},
{\bf [150]} \cite{2009ApJ...698..502B},
{\bf [151]} \cite{2006MNRAS.369.1965C},
{\bf [152]} \cite{birdIV},
{\bf [153]} \cite{2006ATel..957....1M},
{\bf [154]} \cite{2009A&A...502..787Z},
{\bf [155]} \cite{2007ApJ...659..549C},
{\bf [156]} \cite{2010MNRAS.402.1161P},
{\bf [157]} \cite{2009AstL...35...33R},
{\bf [158]} \cite{1995ApJS..100...69F},
{\bf [159]} \cite{2002ApJS..141...23E},
{\bf [160]} \cite{2008ATel.1425....1M},
{\bf [161]} \cite{2009arXiv0912.1519L},
{\bf [162]} \cite{2008ATel.1686....1M},
{\bf [163]} \cite{2009arXiv0910.5603T},
{\bf [164]} \cite{krietal07b},
{\bf [165]} \cite{2009ATel.2257....1E},
{\bf [166]} \cite{2009ATel.2193....1R},
{\bf [167]} \cite{2009A&A...493..893L},
{\bf [168]} \cite{2010arXiv1001.0568M},
{\bf [169]} \cite{2007ApJ...657L.109P},
{\bf [170]} \cite{2009A&A...501.1031K},
{\bf [171]} \cite{2008A&A...485..195D},
{\bf [172]} \cite{2010ATel.2457....1K},
{\bf [173]} \cite{2010arXiv1002.1076M},
{\bf [174]} \cite{2008AIPC.1085..312T},
{\bf [175]} \cite{2007ATel.1054....1B},
{\bf [176]} \cite{2007ATel.1094....1C},
{\bf [177]} \cite{2007AstL...33..807C},
{\bf [178]} \cite{2007ESASP.622..445C},
{\bf [179]} \cite{2004ATel..350....1G},
{\bf [180]} \cite{2005AstL...31..672G},
{\bf [181]} \cite{2004ATel..266....1G},
{\bf [182]} \cite{2004ATel..332....1G},
{\bf [183]} \cite{2007AstL...33..149G},
{\bf [184]} \cite{2005ATel..647....1G},
{\bf [185]} \cite{2007ESASP.622..373G},
{\bf [186]} \cite{1991SvAL...17...42S},
{\bf [187]} \cite{2003A&A...411L.357M},
{\bf [188]} \cite{KarasevOtranto},
{\bf [189]} \cite{2005ATel..446....1G},
{\bf [190]} \cite{ 2010arXiv1004.4086C},
{\bf [191]} \cite{grebenev2010},
{\bf [192]} \cite{2010arXiv1003.3741R},
{\bf [193]} \cite{2008A&A...484..783C},
{\bf [194]} \cite{2009arXiv0910.3074R},
{\bf [195]} \cite{2010arXiv1006.1272M}.

\normalsize

%\bibliography{ridge,integral}

\include{table}

\end{document}

%% file: table.tex
\tabletypesize{\scriptsize}
\begin{deluxetable}{rlrrcccl}
\tablecaption{The catalog of sources detected during the INTEGRAL/IBIS 7-year all-sky survey. For the column descriptions see Sect.\ref{sec:catalog}. \label{tab:catalog}}
\tablewidth{0pt}
\tablehead{
\colhead{Id} &
\colhead{Name} &
\colhead{RA\tablenotemark{\dag}} &
\colhead{Dec\tablenotemark{\dag}} &
\colhead{$F_{\rm 17-60~keV}^{\star}$} &
\colhead{Type} &
\colhead{Ref.\tablenotemark{\star\star}} &
\colhead{Notes\tablenotemark{\star\star\star}} \cr
\colhead{ } &
\colhead{ } &
\colhead{(deg)} &
\colhead{(deg)} &
\colhead{$\textrm{erg~} \textrm{cm}^{-2}\textrm{s}^{-1}$} &
\colhead{ } &
\colhead{ } &
\colhead{} 
}
\startdata
1 & IGR J00040+7020 &    1.008 &   70.322  & $0.87\pm0.13$ & AGN & 153,116 & Sy2 z=0.096; \\
2 & IGR J00234+6141 &    5.723 &   61.700  & $0.79\pm0.10$ & CV & 43,50 &  \\
3 & TYCHO SNR &    6.334 &   64.150  & $0.92\pm0.10$ & SNR &  &  \\
4 & SWIFT J0025.8+6818 &    6.387 &   68.362  & $1.01\pm0.12$ & AGN &  & Sy2 z=0.012; \\
5 & V709 Cas &    7.205 &   59.300  & $5.85\pm0.11$ & CV & 36 &  \\
6 & IGR J00291+5934 &    7.254 &   59.563  & $31.29\pm0.32$\tablenotemark{R261:264} & ~~~~~~LMXB~~~~~~ &  &  \\
7 & IGR J00335+6126 &    8.397 &   61.448  & $0.67\pm0.10$ & AGN & 149,134 & Sy1.5 z=0.105; =IGR J00333+6122 \\
8 & 87GB003300.9+593328 &    8.977 &   59.827  & $1.40\pm0.11$ & AGN & 1 & Blazar z=0.086; \\
\enddata
\tablenotetext{\dag}{The positional accuracy of sources detected by IBIS/ISGRI depends on the source significance \citep{gros2003}. The estimated $68\%$ confidence intervals for sources detected at 5--6, 10, and $>20\sigma$ are $2.1$\arcmin, $1.5$\arcmin, and $<0.8$\arcmin,  respectively (K07).}
\tablenotetext{\star}{Source flux was measured on the 7-year time-averaged map or over the indicated spacecraft revolution interval RXXX:XXX. Flux is expressed in units of $10^{-11}\textrm{erg}$~$\textrm{cm}^{-2}\textrm{s}^{-1}$.}
\tablenotetext{\star\star}{The number denotes reference listed in Sect.\ref{section:references}.}
\tablenotetext{\star\star\star}{The spatial confusion with the source XXX is indicated as CXXX. The measured flux of sources being in spatial confusion should be taken with the caution.}
\end{deluxetable}